\documentclass[nofootinbib,amsmath,twocolumn,notitlepage,preprintnumbers,floatfix]{revtex4-2}
\usepackage{multirow}
\usepackage{amscd, amsfonts, amsmath, amssymb, amsthm, eso-pic, esvect,  graphicx, latexsym, physics}
\usepackage[x11names]{xcolor}
\usepackage{hyperref}
\usepackage{graphicx}
\usepackage{subcaption}
\usepackage{bm}
\usepackage{float}
\usepackage{mathrsfs}
\usepackage{colortbl}
\usepackage{siunitx} 
\usepackage{subfiles}
\usepackage[english]{babel}
\usepackage[justification=raggedright,singlelinecheck=false]{caption}
\usepackage{cleveref}
\usepackage{mathtools}
\usepackage[normalem]{ulem}
\usepackage{cancel}
\usepackage{lipsum}
\usepackage{ragged2e}

\providecommand{\refstepcounter}[1]{}   
\providecommand{\theequation}{\#}       
\newcommand{\atag}{\refstepcounter{equation}\tag{\theequation}}
\newcounter{num}

\newcommand{\edit}[1]{#1}

\usepackage{hyperref}

\usepackage[english]{babel}

\begin{document}

\title{Reappraisal of the Constraints on Heavy Axion-like Particles from Gamma-Ray Bursts}

\author{Christopher V. Cappiello$^{a}$}
\thanks{cappiello@wustl.edu}
\author{Saurav Das$^a$}
\thanks{s.das@wustl.edu} 
\author{P.~S.~Bhupal Dev$^{a,b}$}
\thanks{bdev@wustl.edu}
\author{Takuya Okawa$^{c,d,e}$}
\thanks{tokawa@sissa.it}
\author{Soebur Razzaque$^{f,g,h}$}
\thanks{srazzaque@uj.ac.za}

\medskip

\affiliation{$^a$Department of Physics and McDonnell Center for the Space Sciences, Washington University, Saint Louis, MO 63130, USA}
\affiliation{$^b$PRISMA$^{++}$ Cluster of Excellence \& Mainz Institute for Theoretical Physics, 
Johannes Gutenberg-Universit\"{a}t Mainz, 55099 Mainz, Germany}
\affiliation{$^c$SISSA, International School for Advanced Studies, via Bonomea 265, 34136 Trieste, Italy}
\affiliation{$^d$IFPU, Institute for Fundamental Physics of the Universe, via Beirut 2, 34014 Trieste, Italy}
\affiliation{$^e$INFN, Sezione di Trieste, via Valerio 2, 34127 Trieste, Italy}
\affiliation{$^f$Centre for Astro-Particle Physics (CAPP) and Department of Physics, University of Johannesburg, PO Box 524, Auckland Park 2006, South Africa}
\affiliation{$^g$Department of Physics, The George Washington University, Washington, DC 20052, USA}
\affiliation{$^h$National Institute for Theoretical and Computational Sciences (NITheCS), Private Bag X1, Matieland, South Africa}


\begin{abstract}
   We reassess existing limits and derive new constraints on heavy axion-like particle (ALP) coupling to photons using gamma-ray bursts (GRBs). ALPs can be produced in the hot dense fireball plasma during the initial stage of GRB outflow, thus potentially disrupting the primary fireball and altering the GRB luminosity. We consider the ALP production rate for various GRB parameters in two different energy injection scenarios of GRB fireball formation, and point out that ALP production is less efficient than previously assumed unless a GRB event is exceptionally energetic. 
   We update the existing energy loss bounds using more realistic GRB parameters. We also point out that in the region of parameter space previously constrained by GRB luminosity  criterion, ALP production turns out to be still efficient enough to form a secondary fireball via ALP decay to two photons and their subsequent annihilation to electron-positron pair. This secondary fireball  
   reprocesses the gamma-rays from heavy ALP decay into $X$-rays, emitted isotropically from its surface, thus allowing us to probe $\mathcal{O}(100~\mathrm{MeV})$-scale ALPs indirectly using $X$-ray (or future MeV gamma-ray) telescopes, not necessarily directed toward the GRB jet itself. We show that the future point-source sensitivity of $X$-ray and MeV gamma-ray telescopes may allow us to constrain new ALP parameter space.
\end{abstract}

\maketitle

\section{Introduction}
\label{sec: Introduction}


Axion-like particles (ALPs) are generalizations of the QCD axion~\cite{Weinberg:1977ma,Wilczek:1977pj} and often appear naturally in string theory constructions~\cite{Svrcek:2006yi,Arvanitaki:2009fg}. Unlike the QCD axion, they are not necessarily tied to solving the strong CP problem~\cite{Peccei:1977hh,Peccei:1977ur}, thus allowing their mass and coupling strengths to be much more flexible. Because of their potential accessibility, ALPs have motivated a wide range of terrestrial, astrophysical, cosmological searches; see {\it e.g.}, Refs.~\cite{Irastorza:2018dyq, Choi:2020rgn,  Arza:2026rsl}. 

ALPs arise as pseudo-Nambu-Goldstone bosons of a broken global symmetry (a shift symmetry). That symmetry dictates how they interact with the Standard Model (SM) particles. Direct couplings, {\it e.g.},~to SM fermions, are often derivative-suppressed. The most important low-energy interactions come from anomalies, which lead to couplings to gauge bosons. In particular, a coupling to photons of the form  $g_{a\gamma} a F_{\mu\nu}\widetilde{F}^{\mu\nu}$ (where $a$ is the ALP field, $F_{\mu\nu}$ is the electromagnetic field strength tensor and $\widetilde{F}^{\mu\nu}$ its dual) is well-motivated because it arises naturally from triangle anomalies, survives at low energies after electroweak symmetry breaking and leads to clean experimental signatures.

\begin{figure}[b!]
    \centering
    \includegraphics[width=\linewidth]{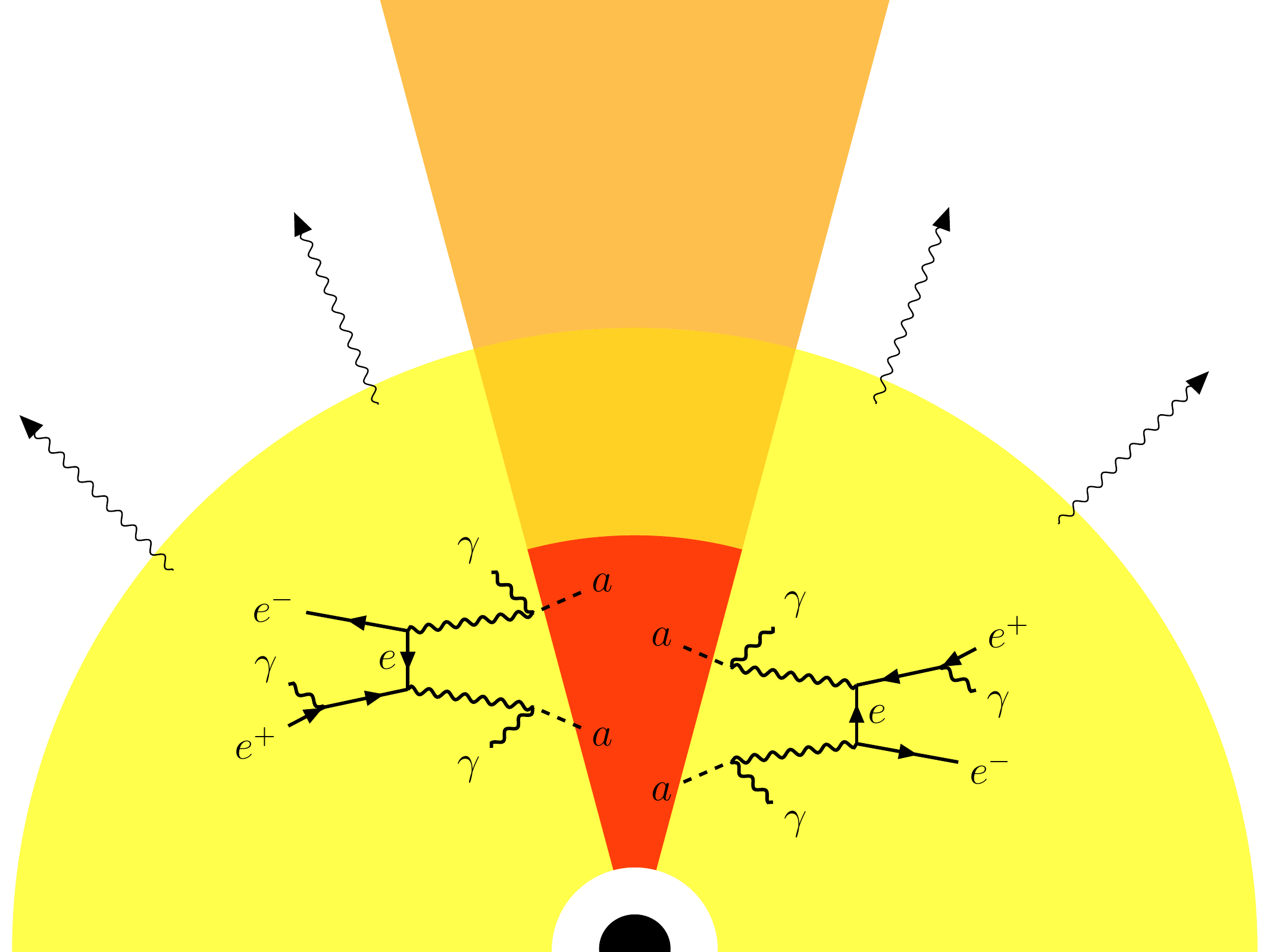}
    \captionsetup{justification=Justified}
    \caption{Production of a secondary fireball (yellow) by axions produced in the primary, Standard Model fireball (red). Also shown is the GRB jet (orange) and the progenitor (black blob) at the center.}
    \label{fig:fireball}
\end{figure}

Among the various methods of searching for and constraining ALPs, the most relevant to the present work is ALP production via its coupling to photons in astrophysical bodies such as stars and supernovae~\cite{Caputo:2024oqc,Carenza:2024ehj}. In the presence of a hot and dense plasma as in the stellar cores, ALPs are copiously produced via the Primakoff process ($\gamma + e \to a + e$) and/or the photon coalescence ($\gamma + \gamma \to a$). This can lead to various observable signals such as (i) the anomalous cooling~\cite{Raffelt:1996wa} relative to that due to neutrino/photon emission, and (ii)  $\gamma$-ray signal arising from either ALP conversion to photons in presence of an external magnetic field (for lighter ALPs)~\cite{Mikheev:1998bg,Caputo:2020quz, Carenza:2021alz, Calore:2023srn, Kanodia:2026hze, Fiorillo:2026rgo} or the spontaneous decay of ALPs to photons (for heavier ALPs) produced in the plasma~\cite{Giannotti:2010ty,Jaeckel:2017tud,Hoof:2022xbe,Muller:2023vjm}. Typical stellar core temperatures amount to $\mathcal{O}(10~\mathrm{keV})$ at most, allowing the production of ALPs only up to a few tens of keV in mass~\cite{Ayala:2014pea,Carenza:2020zil,Dessert:2020lil, Lucente:2022wai,Candon:2024eah,Buckley:2024ldr}. On the other hand, more extreme environments like supernovae~\cite{Caputo:2022mah, Ferreira:2022xlw,Muller:2023vjm, Benabou:2024jlj,Fiorillo:2025yzf} and neutron star mergers~\cite{Dev:2023hax, Diamond:2023cto} can produce ALPs up to a few hundred MeV.

Recently, it was proposed~\cite{Ghosh:2025rjh} that even heavier ALPs with mass up to the GeV scale could be produced in gamma ray bursts (GRBs) --  the most energetic explosion events in the Universe~\cite{Peer:2024xlr}.\footnote{See Refs.~\cite{Berezhiani:1999qh,Gianfagna:2004je, Mena:2011xj,Tu:2015lwv,Reynoso:2017mal} for earlier works on axion emission from GRBs.} In particular, it was argued that fireballs that develop in the initial stage of GRB outflow could be as hot as $\mathcal{O}(100~\mathrm{MeV})$. Consequently, a considerable number of ALPs can be produced in the GRB fireball plasma that carry away enough energy to disrupt the fireball and visibly dim the GRB, thus allowing bright GRB observations to put new astrophysical constraints on heavy ALPs up to 5 GeV and ALP-photon coupling down to $g_{a\gamma}\sim 4\times 10^{-12}~{\rm GeV}^{-1}$~\cite{Ghosh:2025rjh}.  

In this work, we closely examine the temperature profile of the GRB fireball based on two contrasting assumptions: (i) GRB fireballs are fueled by a relativistic outflow over a timescale comparable to a duration of the burst and (ii) the entire GRB energy is injected into a region of radius slightly larger than the Schwarzschild radius of the remnant object that launches a relativistic jet. We point out that in both scenarios, only anomalously energetic GRB events can sustain a burst with $\mathcal{O}(100~\mathrm{MeV})$ temperature, which is rather unrealistic. For physically realistic situations, we find that the ALP production inside the GRB fireball is far less efficient, and the cooling bound is significantly weaker than that reported in Ref.~\cite{Ghosh:2025rjh}. We assess the cooling bounds for several choices of GRB energy and luminosity values. 

We also find that similar to the supernova~\cite{Diamond:2023scc} and neutron star merger~\cite{Diamond:2023cto} cases, a secondary fireball could be formed by photons from the decay of ALPs ($a\to \gamma\gamma$) that are produced inside the jet of the primary GRB fireball; see Fig.~\ref{fig:fireball} for an illustration. This could happen after the ALPs escape the primary GRB fireball, and depending on the values of $g_{a\gamma}$ and the ALP mass $m_a$, would in principle lead to detectable photon signals at $X$-ray telescopes such as the Swift/BAT~\cite{Lien:2025zzo}. We argue that a search for these $X$-rays from the surface of the  secondary fireball is a more effective probe of the ALP emission by GRBs, as it does not require any assumption about what the baseline luminosity of a GRB should be in the absence of anomalous cooling. We also point out that such an $X$-ray signal is expected even if the jet is not directed toward us since the secondary fireball emits $X$-rays nearly isotropically regardless of the direction of the GRB jet. 
This analysis is conceptually similar to that performed in Ref.~\cite{Diamond:2023cto} for neutron star mergers, but uses ALPs sourced from the primary GRB fireball rather than from the hypermassive neutron star itself. While this leads to somewhat weaker bounds than those in  Ref.~\cite{Diamond:2023cto}, it does not rely on any assumption about the lifetime of the hypermassive neutron star merger remnant, and applies even in the case of immediate collapse of the merger remnant to a black hole.

The rest of the paper is organized as follows: In Sec.~\ref{sec:parameters}, we discuss the typical GRB parameters we will be using in our analysis. In Sec.~\ref{sec:alps production}, we consider the ALP production in the primary GRB fireball. Sec.~\ref{sec:secondary} discusses the secondary fireball formation from the ALP-induced photons. In Sec.~\ref{sec:diffuse bound}, we present our main results. Our conclusions are given in Sec.~\ref{sec: conclusions}. In Appendix~\ref{sec:previous}, we make some comments on the validity of the previous results from Ref.~\cite{Ghosh:2025rjh}. 

\section{GRB Parameters}
\label{sec:parameters}
GRBs are extremely powerful transient events of extragalactic origin~\cite{Gehrels:2013xd, Luongo:2021pjs, Vigliano:2024key}. Over the last few decades, improvements in the angular localizations~\cite{1995SPIE.2517..169P, Costa:1997obd} have allowed us to map the GRBs to their progenitors. It is believed that GRBs are powered by a relativistic jet that is launched by gravitational capture of surrounding material by a newly formed compact object, typically a black hole.  The observation of GW170817 by LIGO~\cite{LIGOScientific:2017vwq} and the corresponding GRB event observed by Fermi-GBM and INTEGRAL~\cite{LIGOScientific:2017zic} confirmed a binary neutron star merger as the progenitor of a class of GRBs. The diversity of GRB events have led to classify the GRBs using the most important of their observational characteristics. The great observational and morphological complexity of the GRBs calls for GRB classification based upon various parameters like  GRB progenitor types or hardness of the GRB spectrum.  However, the primary classification of GRBs is based upon the temporal length of prompt gamma-ray emission produced by them.  

\subsection{Long versus short GRBs}
\label{sec:longshort}

The duration of GRBs has been observed to have a wide range, starting from $\sim 10~\text{ms}$ up to $\sim 2000~\text{s}$~\cite{Vigliano:2024key}. In addition to the variation in temporal length, the emission spectrum also shows significant differences. The peak spectrum shifts towards `harder' for GRBs lasting for shorter durations. To classify GRBs based on their temporal length, duration measures such as $T_{90}~(T_{50})$ are used, defined as the time frame in which $90\%~(50\%)$ of the background-subtracted photon counts, namely, from $5\%$ to $95\%$ ($25\%$ to $75\%$) of the total photons, are observed. Based on this temporal measure, it was observed that GRB catalogs follow a bimodal distribution~\cite{Mazets:1981sye, Kouveliotou:1993yx, 2012grb..book.....K}, hence can be divided into two classes, the short/hard GRB (sGRB) for $T_{90}< 2~ \text{s}$ and long/soft GRB (lGRB) for $T_{90} > 2~\text{s}$.

However, this type of classification can run into several problems. Since GRBs emit radiation over a wide range of frequencies and emission times of different frequencies do not coincide, with softer frequencies emitted later, measures like $T_{90}~(T_{50})$ are frequency-dependent. Apart from that, the existence of ultra-long GRBs, characterized by the observation of prompt emission for over $10^4~\text{s}$ begs the question of whether it is a distinct class of GRBs or the long tail of the lGRBs. Furthermore, cosmological events like $X$-ray flashes~\cite{Heise:2001yh, Piran:2012qg} that show similar features of GRB emission but in a lower energy band are  suspected to have the same astrophysical properties of GRBs but observed off-axis. 

Due to these issues, a further classification of GRBs based on the progenitor was proposed. Observations support the fact that sGRBs have different locales in contrast to lGRBs.  
The sGRBs are predominantly at the edge of irregular galaxies, away from the star forming regions, whereas lGRBs are observed in the star forming region of the star forming host galaxies~\cite{Bloom:2005qx, Gehrels:2005qk, Barthelmy:2005bx,vandenHeuvel:2013bka}. These differences have lent support to the assumption that vastly different mechanisms are responsible for short and long GRBs. Long GRBs/Type II GRBs are believed to be associated with the death of massive stars~\cite{Wang:1998yu}. On the other hand, short/Type I GRBs are thought to be due to coalescence of compact objects like binary neutron stars mergers or black hole-neutron star  mergers~\cite{Berger:2013jza}. We have checked that the sGRBs typically give us better ALP constraints than lGRBs, mainly due to their higher temperature, and therefore, we only consider sGRBs in our subsequent analysis. 
\subsection{Opening angle and isotropic-equivalent energy}
\label{sec:energy}
The total isotropic energy output\footnote{ Isotropic energy output is defined as the energy budget inferred from observations under the assumption that the emission is isotropic.} for GRBs is estimated to be in the range of $10^{50}-10^{54}~\text{ergs}$; see Appendix~\ref{sec:previous} for details. The huge isotropic energy released means ${\cal O}(1-3) M_\odot$, where $M_\odot$ is the solar mass, is converted into radiation in a few seconds.\footnote{Long GRBs can have a higher progenitor mass up to $10M_\odot$ and correspondingly higher total isotropic energy output up to $10^{55}$ ergs.} This energy output and efficiency of energy conversion puts severe constraints on the possible mechanisms involved in GRBs. The short variability time scale of the emission as well as non-thermal nature of the spectra alluded to possible role of relativistic jets at play~\cite{Goodman:1986az, Piran:2004ba, 2012grb..book.....K}. The jets also substantially alleviate the requirement of huge energy output by decreasing it by an angular factor: 
\begin{align}
    E_{\gamma}^\mathrm{tot} = E_{\text{iso}} (1-\cos\theta) \, , 
\end{align}
where $E_{\gamma}^\mathrm{tot}$ and $E_{\text{iso}}$ denote total energy of emitted radiation and the inferred total isotropic radiation respectively, and $\theta$ denotes the opening angle of the jet. The jet can be thought of as a highly relativistic flow of plasma along a doubly-coned structure with an opening angle $\theta$. 

In general, it is difficult to estimate the jet opening angle due to the challenges in measurement of the viewing angle of the observer with respect to the jet axis \cite{Goldstein:2015fib}. However, a few important features of the relevant astrophysical process like `jet breaks' can be obtained from observing the spectra. The dynamics of the jet that best explain the features of the spectra are as follows: At the beginning, the jet is narrowly beamed and the bulk Lorentz factor $\Gamma$ satisfies $\Gamma^{-1} < \theta$. The prompt emission produced during this phase is only visible to an on-axis observer. However, as the jet expands, the ejecta decelerates to $\Gamma^{-1} > \theta$. The afterglow emission from this phase can be observed both by an on-axis observer, as well as by an off-axis observer. 

The few general properties of GRB that can be inferred from various features of the observed spectra and from the correlation of the observed quantities can be summarized as follows: The burst energy is gravitational in nature and the central engine powering this burst is typically a black hole~\cite{Paczynski:1986px}. A significant fraction of the burst energy is initially converted into kinetic energy of the jet that causes the relativistic outflow. The dissipation of the kinetic energy causes the initial prompt emission and the subsequent afterglow~\cite{Panaitescu:1998zf, Sari:1999mr}.
\subsection{Temperature and Lorentz boost profiles}
\label{sec:temp}
The above discussion points to a highly collimated and highly relativistic outflow of thermalized plasma, originating from around a newly formed compact object. The key GRB parameters like temperature and the Lorentz boost factor of the relativistic plasma can be described in two different reference frames. One is the frame of the central engine, which is the same as the lab frame. But for our purposes, it is more appropriate to describe the relevant parameters in the comoving frame of an infinitesimally thin shell in the jet. Under the assumption of spherical symmetry, the temperature profile and Lorentz factor of the steady-state outflow follow simple scaling relations~\cite{1982MNRAS.199..833F,Paczynski:1986px}: $T\propto r^{-1}$ and $\Gamma\propto r$, {\it i.e.},
\begin{align}
    T(r) = T_{i} \frac{r_i}{r} \, ,  && \Gamma(r) = \frac{r}{r_i} \, , \label{eq:scaling}
\end{align}
where $T_{i}$ is the temperature evaluated at the initial radius $r = r_i$ where the fireball is launched. Following Ref.~\cite{Tu:2015lwv} (which in turn bases its choice on the smallest observationally derived value presented in Ref.~\cite{Peer:2015pzy}), we take it to be three times the gravitational (or Schwarzschild) radius of the remnant, $r_i=3 r_s$, where $r_s = 2GM_r/c^2$, $M_r$ is the mass of the remnant, and $G$ is the gravitational constant.\footnote{We use natural units with $c=1$, but for clarity, keep the explicit $c$-dependence here and in some expressions below.} Assuming a representative value of $M_r = 3 M_\odot$ for the remnant mass, we get $r_s = 8.9 \times 10^5~\mathrm{cm}$. Throughout this work, we take $r_i = 3r_s$, which is consistent with Ref.~\cite{Tu:2015lwv}, given the value of $r_s$ quoted above for a 3$M_{\odot}$ remnant. 

The temperature $T_i$ of the fireball at the jet launching radius  can be calculated under two different assumptions. The more commonly assumed scenario is that the GRBs are powered by a relativistic outflow lasting approximately the duration of the burst~\cite{Rees:1994nw, Daigne:2002zg}. In this picture, the temperature $T_i$ is given in terms of the isotropic-equivalent  luminosity $L$ by~\cite{Bahcall:2000sa,Rossi:2005uc,Razzaque:2006ju}
\begin{equation}\label{eq:luminositytemperature}
    T_i \simeq (2.1~\text{MeV}) \times \left[\frac{L}{10^{52} \, \textrm{erg/s}}\left(\frac{10^{6.5} \, \textrm{cm}}{r_i}\right)^2\right]^{1/4}.
\end{equation}
This formula results in a conservative estimate of the temperature for a given energy output, as we will discuss next. 

Another possibility is to assume that all of the energy from the GRB is deposited instantaneously inside a region that is only slightly larger than the horizon of the progenitor mass~\cite{Koers:2005ya}. If the plasma consisting of electrons, positrons, baryons and radiation is optically thick to radiation, it can only cool by adiabatic expansion. The initial fireball temperature $T_i$ in this case can be obtained using the fact that in thermodynamic equilibrium, the energy density and temperature are related by 
\begin{align}
 \frac{E}{V} = \frac{\pi^2}{30}g_*T^4 \, ,
\end{align}
where $E$ is the energy released from the source of the fireball,
$V$ is the fireball volume
and $g_*$ is the total number of effective massless degrees of freedom in the plasma. In the initial fireball phase, photons, electrons, positrons, as well as all three flavors of neutrinos are in thermal equilibrium, so $g_* = 2+\frac{7}{8}(2+2+3\cdot 1+3\cdot 1) = 43/4$. Assuming that all of the GRB's energy is deposited instantaneously in a spherical region of radius $r_i$, the initial temperature $T_i$ can be expressed as~\cite{Tu:2015lwv,Koers:2005ya}
\begin{equation}
\label{eq: Temp_instanteneous}
    T_i^E \simeq (18.2~\text{MeV}) \times \left[\frac{10}{g_*} \frac{E}{10^{52} \, \textrm{erg}} \left(\frac{10^{6.5} \, \textrm{cm}}{r_i}\right)^3\right]^{1/4}.
\end{equation}
This assumption of instantaneous energy injection can lead to much higher temperatures. Using the above formula, and assuming $r_i = r_s$, Ref.~\cite{Ghosh:2025rjh} computes temperatures of several hundred MeV for sGRBs with various remnant masses. We find that an isotropic-equivalent energy $E$ of order $10^{55}$ erg is required to produce these temperatures; for concreteness, $10^{55}$ erg leads to a temperature of 261 MeV from Eq.~\eqref{eq: Temp_instanteneous}. Refs.~\cite{Tu:2015lwv,Koers:2005ya} compute a temperature of 18 MeV using a more conservative value of $E$.
The isotropic-equivalent energy emitted in a GRB roughly varies between $10^{50}-10^{54}$ erg~\cite{Kumar:1999cv,Ghirlanda:2009de, Kumar:2014upa, Wanderman:2009es}. Using Eqs.~\eqref{eq:luminositytemperature} and ~\eqref{eq: Temp_instanteneous}, we tabulate some benchmark values of the initial fireball temperature for different values of the energy release in Table~\ref{tab:GRB Parameters} in the Appendix and recalculate the cooling bounds in the ALP parameter space (see Fig.~\ref{fig:energylosscontours}). 



\section{ALP production in GRB fireball}
\label{sec:alps production}
Here we consider the interaction of ALPs with the electromagnetic field, encoded by the Lagrangian 
\begin{align}
    \mathcal{L}_a \supset \frac{1}{2}(\partial_\mu a)^2-\frac{1}{2} m_a^2 a^2 + \frac{g_{a\gamma}}{4}a F_{\mu\nu} \widetilde{F}^{\mu\nu} \, ,
\end{align}
where $F_{\mu\nu}$ is the electromagnetic field strength tensor, $\widetilde{F}^{\mu\nu} \equiv \epsilon^{\mu\nu\rho\sigma} F_{\rho\sigma}/2$ is its dual and $g_{a\gamma}$ is the dimensionful ALP-photon coupling. We are agnostic about ALP couplings to other SM particles, as is the standard practice for such phenomenological studies. ALPs dominantly coupling to photons can be motivated from effective field theories~\cite{Brivio:2017ije, Bauer:2017ris, Bauer:2020jbp} and in ultraviolet-complete models, {\it e.g.}, by engineering anomaly coefficients (no color anomaly) and restricting to electroweak gauge sectors as in KSVZ-type models~\cite{Kim:2008hd, DiLuzio:2016sbl}, or by using string-theoretic constructions where ALPs selectively couple to $U(1)$ fields~\cite{Arvanitaki:2009fg}.

Due to the coupling to the photon field, ALPs can be produced via either the Primakoff process ($\gamma + Ze \to a + Ze$) or the photon coalescence process ($\gamma + \gamma \to a$) in a thermal plasma. For relatively heavier ALPs ($m_a\gtrsim 100$ MeV), the photon coalescence turns out to be the dominant process~\cite{Carenza:2020zil,
Caputo:2022mah}; however, we include the Primakoff contribution in our calculation for completeness.

\paragraph{Photon coalescence:}
The invariant squared amplitude for the photon coalescence process is given by 
\begin{align}
    \label{eq:invariant matrix}
    \sum |\mathcal{M}_{\gamma\gamma \rightarrow a}|^2 = \frac{1}{2} g_{a\gamma}^2\, m_a^2 \left( m_a^2 - 4 \omega_\mathrm{pl}^2 \right) \, ,
\end{align}
where $\omega_{\rm pl}$ is the plasma frequency or the effective photon mass in the plasma. In the relativistic limit ($T \gg m_e$), it is given by~\cite{Braaten:1993jw}
\begin{align}
    \omega_{\rm pl}^2=\frac{4\alpha}{3\pi}\left(\mu_e^2+\frac{1}{3}\pi^2T^2\right) \, ,
\end{align} 
where $\alpha=e^2/4\pi\simeq 1/137$ is the fine-structure constant and $\mu_e$ is the electron chemical potential. In a fireball with electrons, positrons and photons in thermal equilibrium, particle number-changing interactions are possible, so the chemical potential term can be neglected~\cite{Koers:2005ya}. Therefore, 
\begin{align}
    \omega_\mathrm{pl}^2 \simeq  \frac{4 \pi \alpha}{9} T^2 \ \ \ {\rm or}, \ \ \ \omega_\mathrm{pl} \simeq \frac{T}{10} \, .
\end{align}
The ALP production spectrum ({\it i.e.},~number of ALPs produced per unit energy and volume) in the rest frame of the plasma is given by integrating Eq.~\eqref{eq:invariant matrix} over the momentum space. The result is~\cite{Ferreira:2022xlw, Caputo:2022mah, Buckley:2024ldr}
\begin{align*}
\label{eq:alpspectrum}
    \frac{\mathrm{d} \dot{n}_a}{\mathrm{d} E_a}(T)&  = 
    \frac{g_{a\gamma}^2 m_a^2}{128\pi^3} \left(m_a^2 - 4\omega_{\text{pl}}^2\right)
    \frac{T}{e^{E_a/T} - 1} \Theta(m_a-2\omega_{\rm pl})\\
    &  \times \log \left[
    \frac{
    (1 - e^{E_{\mathrm{max}}/T})(e^{E_a/T} - e^{E_{\mathrm{min}}/T})
    }{
    (1 - e^{E_{\mathrm{min}}/T})(e^{E_a/T} - e^{E_{\mathrm{max}}/T})
    }
    \right]
     \atag,
\end{align*}
where $n_a$ is the number density of ALPs produced with energy $E_a$, $\dot{n}_a\equiv \mathrm{d}n_a/\mathrm{d}t$ and $E_\mathrm{max~(min)}$ is the maximum (minimum) energy of each photon that can produce an axion with mass $m_a$: 
\begin{align}
    E_{\mathrm{min}, \mathrm{max}} =
    \frac{1}{2} \left(
    E_a \mp \sqrt{E_a^2 - m_a^2}
    \sqrt{1 - \frac{4 \omega_{\mathrm{pl}}^2}{m_a^2}}
    \right) \, .
\end{align}

\paragraph{Primakoff process:}
While heavy axions ($m_a \gtrsim T$) are mostly produced by the photon coalescence process, the Primakoff process is the dominant production process for lighter axions. The Primakoff production spectrum is given by~\cite{Ferreira:2022xlw, Caputo:2022mah, Buckley:2024ldr}
\begin{align*}
    \label{eqn: the production spectra (Primakoff)}
    &\frac{\dd \dot{n}_{a}}{\dd E_a}(T)
    =\frac{g_{a\gamma}^2 T \kappa^2}{32\pi^3} pk f_B\left(E_{\gamma}\right) \\
    &\times \left\{ \frac{\left[(k+p)^{2}+\kappa^{2}\right]\left[(k-p)^{2}+\kappa^{2}\right]}{4 p k \kappa^{2}} \log \left[\frac{(k+p)^{2}+\kappa^{2}}{(k-p)^{2}+\kappa^{2}}\right] \right. \\
    &\left.\ \ \ \  -\frac{\left(k^{2}-p^{2}\right)^{2}}{4 k p \kappa^{2}} \log \left[\frac{(k+p)^{2}}{(k-p)^{2}}\right]-1 \right\}, \atag
\end{align*}
where $f_{B} = 1/(e^{(E-\mu)/T} - 1)$ is the Bose-Einstein distribution function, $\mathbf{k}$ is the momentum of the incoming photon, $\mathbf{p}$ is the momentum of the outgoing axion, $k$ and $p$ are their modulus values ($|\mathbf{k}| \equiv k$, $|\mathbf{p}| \equiv p$). 
We note that Eq.~\eqref{eqn: the production spectra (Primakoff)} is the approximated expression in the sense that it assumes (i) presence of only two transverse modes with the dispersion relation $\omega^2 = k^2 + \omega_\mathrm{pl}^2$ where $\omega_\mathrm{pl}$ is the plasma frequency, (ii) no recoil of a scatterer, and (iii) negligible Bose enhancement. 
 Since we assume no recoil during the interaction, the energy conservation law expects the same energy of the incoming photon and the outgoing axion. We have taken into account the plasma screening effect by following Ref.~\cite{Raffelt:1985nk}, {\it i.e.}, by replacing the factor $1/|\mathbf{q}|^4$ appearing in $\langle |\mathcal{M}|^2 \rangle$ with $1/(|\mathbf{q}|^2(|\mathbf{q}|^2+\kappa^2))$, where $\kappa$ is the inverse of the Debye screening length, and is given by
\begin{align}
    \label{eqn: screening scale}
    \kappa^{2}=\frac{4 \pi \alpha}{T} n_{e}^{\mathrm{eff}},
\end{align}
for a non-degenerate plasma in the Debye-Huckel approximation and  $n_e^{\mathrm{eff}}$ is the effective number density of electrons for which we use the thermal distribution: 
\begin{align}
    n_e^{\rm eff}(T) = \frac{3}{4}\frac{\zeta(3)}{\pi^2}g_e T^3 \, ,
\end{align}
with $g_e=4$ for the effective degrees of freedom including electrons and positrons. 

After being produced, ALPs can undergo spontaneous decay or be reabsorbed in the plasma via the inverse Primakoff process. The corresponding mean free path is expressed as 
\begin{align}
    \lambda = \left(\lambda_{a\to\gamma\gamma}^{-1} + \lambda_{a\to\gamma}^{-1}\right)^{-1},
\end{align}
where $\lambda_{a \to \gamma\gamma}$ is the mean free path for the spontaneous decay of axions:
\begin{align}
    \lambda_{a \to \gamma\gamma} = \frac{64\pi}{g_{a\gamma}^2 m_a^3} \sqrt{(E_a/m_a)^2 - 1} \, ,
\end{align}
and $\lambda_{a\to \gamma}$ is that for the inverse Primakoff process:
\begin{align}
    \lambda_{a \to \gamma} = \frac{\beta_a^2 E_a^2}{2 \pi^2 e^{E_a/T}} \left(\frac{\mathrm{d} \dot{n}_a}{\mathrm{d} E_a }\right)^{-1},
\end{align}
where $\beta_a = \sqrt{1 - (m_a/E_a)^2}$ is the ALP velocity. 

Outside the fireball plasma, only spontaneous decay is relevant. Thus, the lifetime of ALP can be found as
\begin{align*}
    \tau_a &= \frac{\lambda_{a \to \gamma\gamma}}{\beta}
    \\
    &\simeq (10^{-2}~\mathrm{s})\frac{E_a}{m_a}
   \left( \frac{10^{-10}~\mathrm{GeV}^{-1}}{g_{a\gamma}} \right)^2 \left( \frac{1~\mathrm{GeV}}{m_a} \right)^3 . \atag
\end{align*}
This expression implies that ALPs produced inside the fireball are likely to decay en route to the Earth if they successfully escape from the fireball, thus potentially yielding a large photon flux detectable by gamma-ray telescopes. 

\section{Secondary Fireball Formation}
\label{sec:secondary}
If enough axions are produced from the GRB fireball, and if they decay to photons in a small enough volume, these photons may begin to pair-produce electrons and positrons, ultimately forming a shell of plasma that we refer to as the secondary fireball; see Fig.~\ref{fig:fireball}. The formation of this secondary fireball is dominated by ALPs from the high-temperature region at the core of the initial fireball, resulting in a peculiar geometry where the secondary fireball would surround the base of the initial fireball, but would be pierced by the outer regions of the initial fireball and/or the jet further out. This secondary fireball would produce thermal photons with lower energy than those that would come directly from ALP decays, which might be possible to search for using lower-frequency telescopes. The main advantage is that since the thermal photon emission from the secondary fireball is isotropic, the telescopes need not be directly along the line of sight of the jet to observe this signal. Fireball formation from astrophysically produced ALPs has been treated in detail in Refs.~\cite{Diamond:2023scc,Diamond:2023cto}, and in this Section, we largely follow the procedure from these references.

Two criteria must be met for secondary fireball formation to occur. First, the pair-production rate must be sufficiently high, leading to the requirement~\cite{Diamond:2023scc}

\begin{equation}
\label{Eq: pair production}
    \frac{N}{2\pi r_{\gamma}^2} \sigma_{\gamma \gamma \rightarrow e^+e^-} \gg 1\,.
\end{equation}
Here, $r_{\gamma}$ is the radius of the secondary fireball and $N$ is the total number of ALPs produced. The pair-production cross section is given by~\cite{Diamond:2023scc}

\begin{align}
    \sigma_{\gamma \gamma \rightarrow e^+e^-} &= \frac{\pi \alpha^2}{E_{\gamma}^2} \left[\left(2 + \frac{2m_e^2}{E_{\gamma}^2} - \frac{m_e^4}{E_{\gamma}^4}\right) \log \left|\frac{E_{\gamma}}{m_e} + \sqrt{\frac{E_{\gamma}^2}{m_e^2} - 1} \right| \right. 
 \nonumber \\
    &\qquad\qquad \left. - \sqrt{1 - \frac{m_e^2}{E_{\gamma}^2} \left(1 + \frac{m_e^2}{E_{\gamma}^2}\right)}\, \right]\,. 
\end{align}
For $r_{\gamma}$, we use the average value computed as

\begin{equation}
    \langle r_{\gamma} \rangle = \left\langle c\edit{\beta_a}(E_a)\left( \tau_a(E_a) + \frac{r_{esc}}{c\edit{\beta_a}(E_a)} \right) \right\rangle.
\end{equation}
Assuming a duration of $\edit{t_\mathrm{sGRB}}$ = 1 s for the initial fireball, we obtain $N$ by integrating the ALP production spectrum $\mathrm{d}\dot{n}_a/\mathrm{d}E_a$:
\begin{align}
    N &= \edit{t_\mathrm{sGRB}} \, \pi \Delta \theta^2 \int_{r_i}^{r_c} \mathrm{d}r\int_{m_a}^{\infty}\mathrm{d}E_a \, r^2 \frac{\textrm{d}\dot{n}_a}{\textrm{d}E_a} e^{-\left(\frac{r_{\rm esc}-r}{\lambda_{\edit{a \to \gamma\gamma}}}\right)} \nonumber 
    \\
    &\quad\times \Theta\left(E_a - m_a \left(1 - \frac{2GM}{r} \right)^{1/2}\right)\,.
\end{align}
For later convenience, we also define two similar integrals, the total radial momentum $P$ and the total energy $E$ carried by the ALPs:
\begin{align}
    P &= \edit{t_\mathrm{sGRB}} \, \pi \Delta \theta^2 \int_{r_i}^{r_c} \mathrm{d}r\int_{m_a}^{\infty}\mathrm{d}E_a \sqrt{E_a^2 - m_a^2} \, r^2 \frac{\textrm{d}\dot{n}_a}{\textrm{d}E_a}  \nonumber \\
    &\times e^{-\left(\frac{r_{\rm esc}-r}{\lambda_{\edit{a \to \gamma\gamma}}}\right)} \Theta\left(E_a - m_a \left(1 - \frac{2GM}{r} \right)^{1/2}\right) \,, \\
    E &= \edit{t_\mathrm{sGRB}} \, \pi \Delta \theta^2 \int_{r_i}^{r_c} \mathrm{d}r\int_{m_a}^{\infty}\mathrm{d}E_a \, r^2 \frac{\textrm{d}\dot{n}_a}{\textrm{d}E_a} e^{-\left(\frac{r_{\rm esc}-r}{\lambda_{\edit{a \to \gamma\gamma}}}\right)}
    \nonumber \\
    &\quad\quad\times \Theta\left(E_a - m_a \left(1 - \frac{2GM}{r} \right)^{1/2}\right)\,.
\end{align}
In these integrals, we take $r_c = 1.5\, r_i$ for the critical radius of the fireball where it starts becoming matter-dominated, and $r_{\rm esc} = 2\, r_i$ for the escape radius beyond which ALPs can escape the fireball before decaying. Because we mostly focus on gamma-ray emission that is not along the axis of the primary fireball, ALPs produced near the base of this fireball need only travel, at most, the diameter of the base of the primary fireball in order to escape from the primary fireball, a distance of order $2r_c$. Increasing the value of $r_c$ has almost no effect on the minimum coupling or the maximum mass for which a secondary fireball would form.

The second criterion is that bremsstrahlung must occur quickly enough to drive the system toward chemical equilibrium and significantly increase the particle number. The requirement for this to occur is
\begin{equation}
    \frac{N}{3 \pi \edit{r_\gamma}^2} \sigma_{ee \rightarrow ee\gamma} \gg 1\,.
    \label{eq:cri2}
\end{equation}
The bremsstrahlung cross section is given by~\cite{Diamond:2023scc}
\begin{equation}
    \sigma_{ee \rightarrow ee\gamma} = \frac{8\alpha^3}{m_e^2}\left(\textrm{log}\left(\frac{2T_{s,i}}{e^{\gamma_E}m_e}\right) + \frac{5}{4}\right)\,,
\end{equation}
where $\gamma_E$ is the Euler-Mascheroni constant and $T_{s,i}$ is the initial temperature of the secondary fireball,  given by
\begin{equation}
    T_{s,i} = \frac{P}{8\Gamma_s \beta_s N}\,.
\end{equation}
Here $\Gamma_s$ is the bulk Lorentz factor of the secondary fireball, easily obtainable from the bulk velocity 
\begin{align}
    \beta_s=\frac{2E}{P}-\sqrt{\frac{4E^2}{P^2}-3}\, .
\end{align}

\begin{figure*}[t!]
  \centering
  \includegraphics[width=0.7\columnwidth]{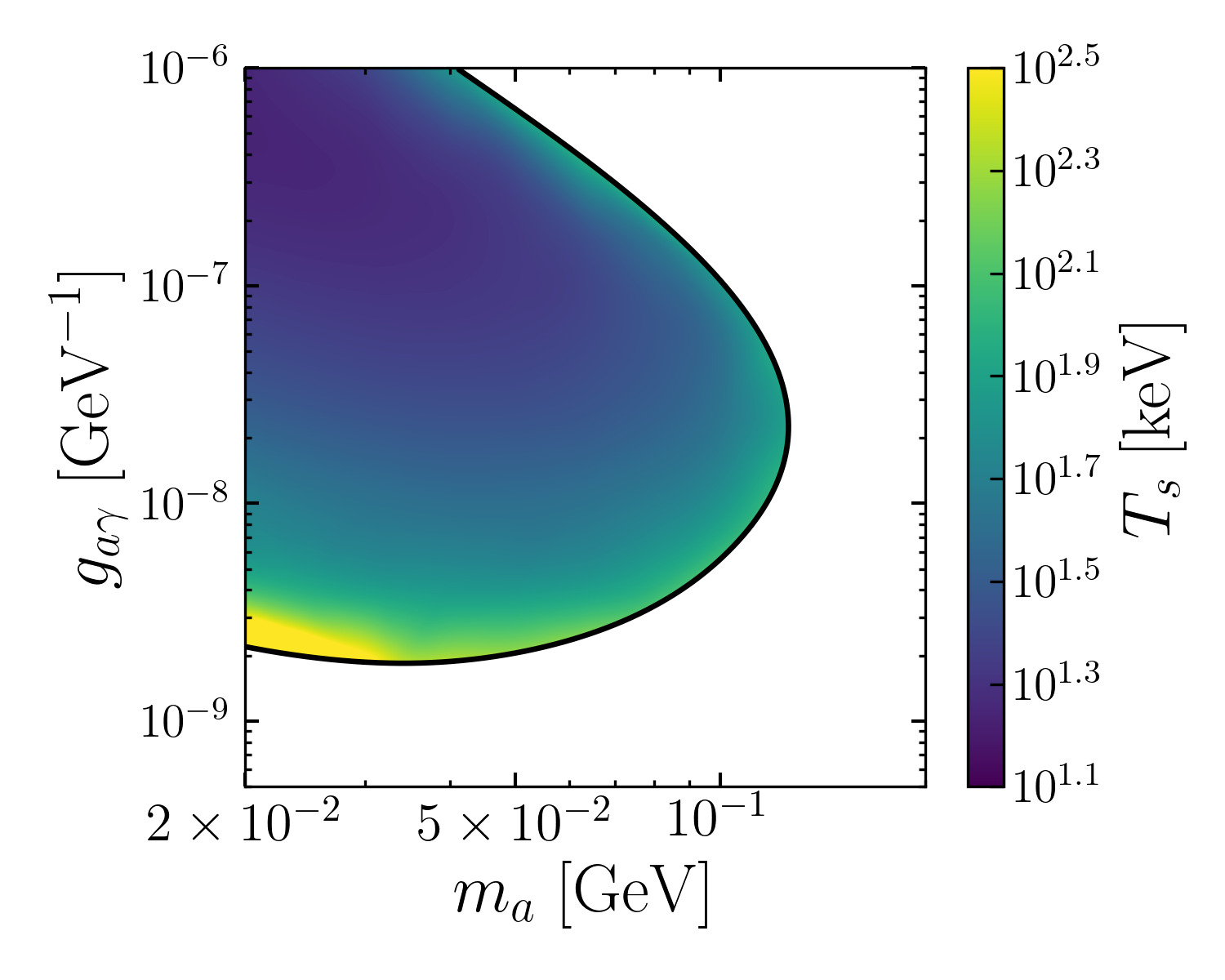}
  \hspace{-4mm}
  \includegraphics[width=0.7\columnwidth]{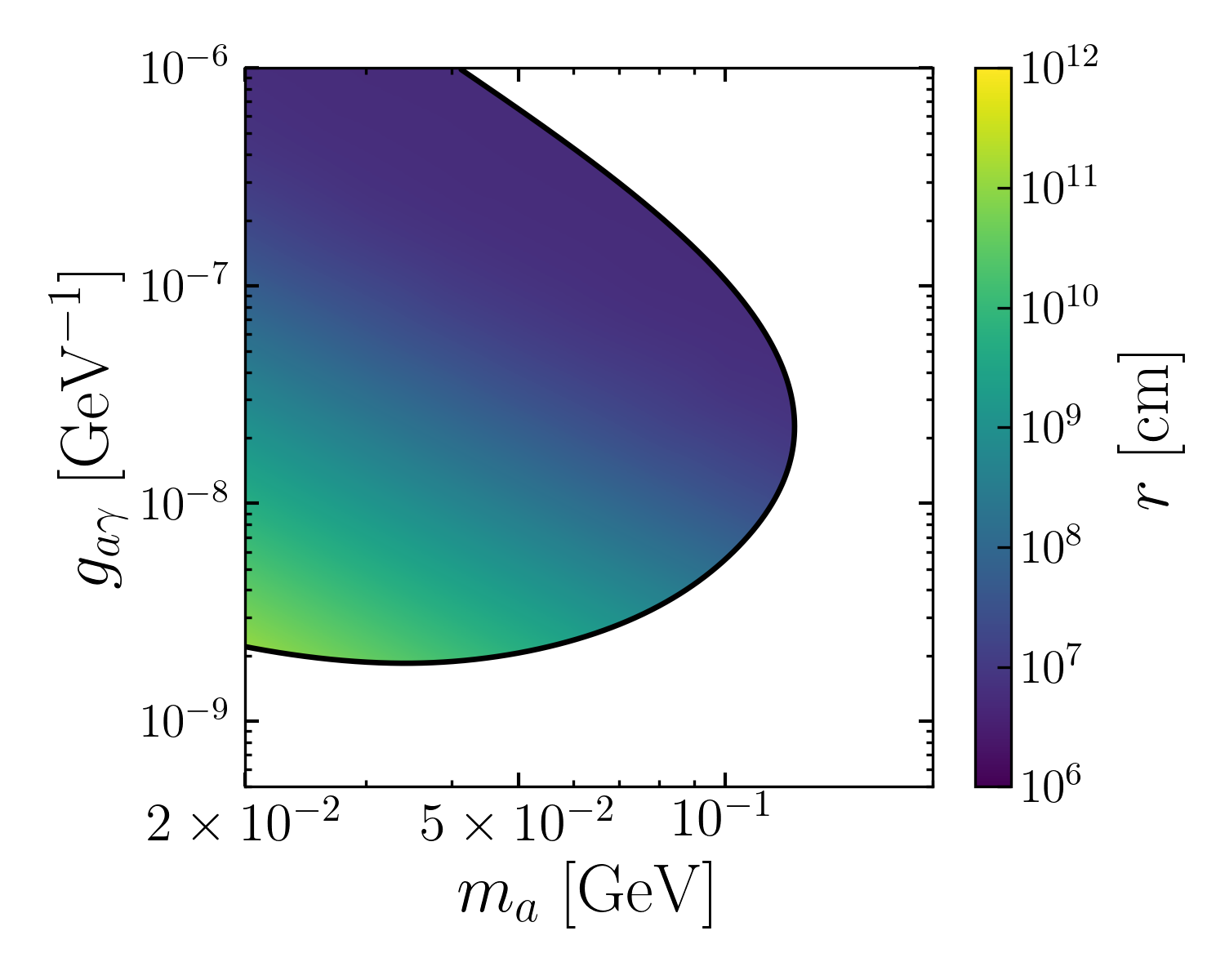}
  \hspace{-4mm}
  \includegraphics[width=0.7\columnwidth]{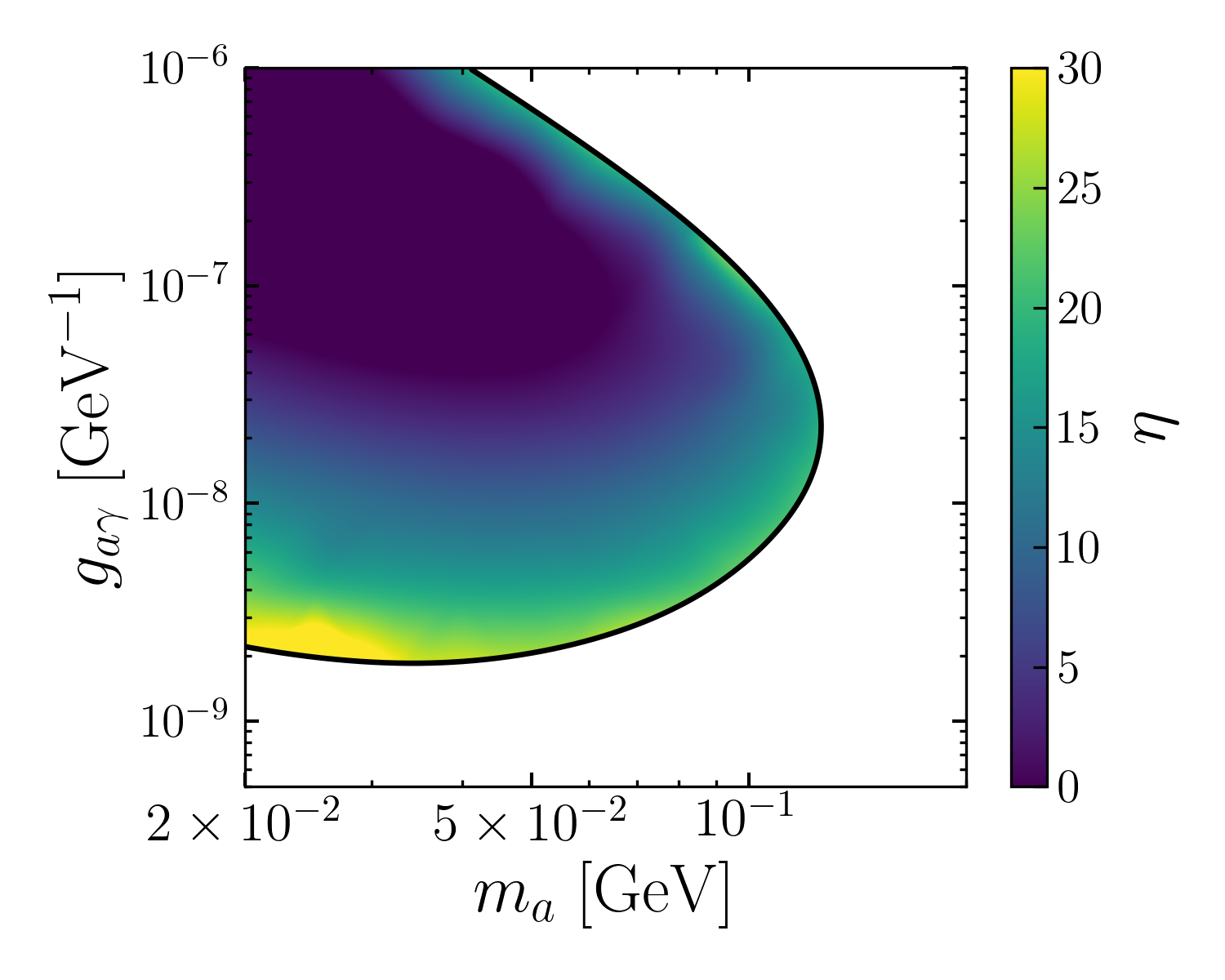}
  \captionsetup{justification=Justified}
  \caption{Secondary fireball properties as functions of the ALP mass and the ALP-photon coupling. The left, middle, and right panels show the secondary fireball temperature on its surface $T_s$, its radius $r$, and its chemical potential (normalized by its temperature) $\eta$, respectively. The secondary fireball is expected to be developed inside the solid black line. }
  \label{fig:fireball temperature}
\end{figure*}

If both of these criteria \eqref{Eq: pair production} and \eqref{eq:cri2} are met, a secondary fireball will form. However, its temperature will not remain at $T_{s,i}$, as rapid particle production decreases the energy per particle. If the secondary fireball never fully thermalizes, which can be the case if thermalization via bremsstrahlung is interrupted by pair annihilation, the final temperature is given approximately by
\begin{equation}
    T_s \sim \dfrac{m_e}{-\textrm{log}\left(\dfrac{3 \pi \Gamma_s \beta_s r^2 m_e^3}{\sqrt{2}\alpha^3 P}\right)}\,,
\end{equation}
and the chemical potential is given by
\begin{equation}
    \mu = T_s\, \textrm{log}\left(\frac{32 \gamma^2 v r^2 \Delta T^4}{\pi P}\right)\,.
\end{equation}
If full thermalization is reached, then $\mu = 0$ and $T = \left(\frac{15E}{4\pi^3 r^2 \Delta}\right)^{1/4}$. In these equations, $\Delta=\sqrt{\langle r_\gamma^2\rangle-\langle r_\gamma\rangle^2}$ is the thickness of the secondary fireball. 

As shown in Fig.~\ref{fig:fireball temperature}, the secondary fireball turns out to be formed in a wide region of the $m_a$--$g_{a\gamma}$ parameter space. The different panels in this figure summarize the secondary fireball properties, namely,  temperature (left), radius (middle), and chemical potential (right).

\begin{figure}
    \centering
    \includegraphics[width=0.99\columnwidth]{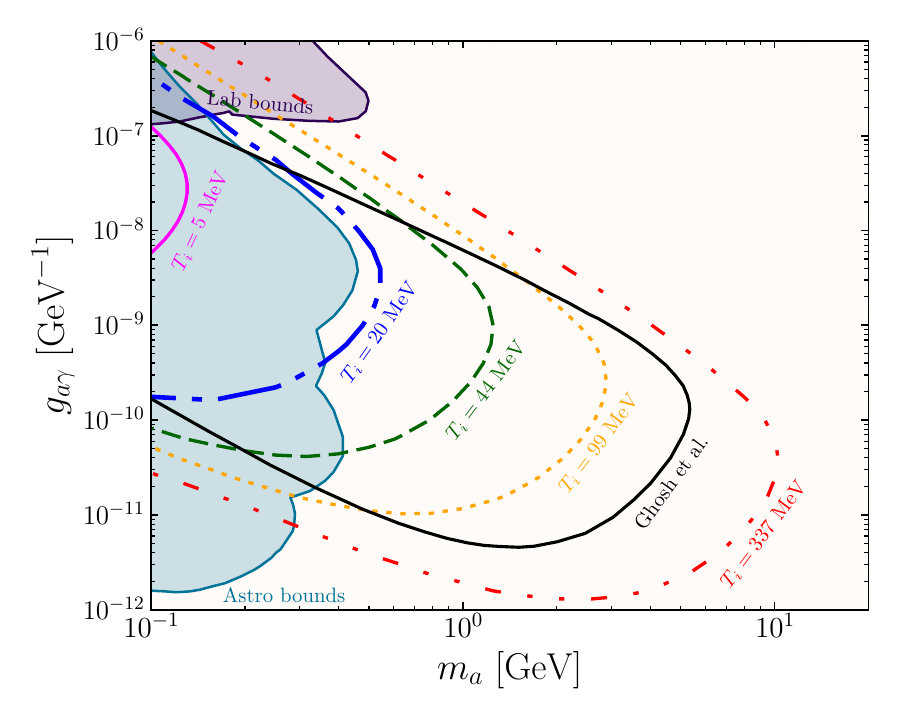}
    \captionsetup{justification=Justified}
    \caption{Secondary fireball formation region for different choices of the initial temperature $T_i$ listed in Table~\ref{tab:GRB Parameters} (as well as for a benchmark temperature of 20 MeV, similar to the temperature assumed in Ref.~\cite{Tu:2015lwv}):  337 MeV (red), 99 MeV (orange), 44 MeV (green), 20 MeV (blue), and 5 MeV (magenta). We also show the current laboratory bounds (purple shaded) and astrophysical constraints (dark cyan shaded). The ALP constraint from Ref.~\cite{Ghosh:2025rjh} is shown by the black contour, but as argued in Appendix~\ref{sec:previous}, may not apply in a realistic GRB scenario. }
    \label{fig:fireball formation}
\end{figure}

Ref.~\cite{Ghosh:2025rjh} claimed that such a secondary fireball could not form in the considered GRB scenario. We disagree with this statement, as detailed below. A simple condition of whether photons can thermalize with each other is  $n_{\gamma} \sigma L \gg 1$, where $n_{\gamma}$ is the photon density, $\sigma$ is the interaction cross section, and $L$ is the size of the region in which the photons are produced. This is in line with the derivation found in Ref.~\cite{Diamond:2023scc}, with $L$ taken to be the standard deviation of the axion decay length distribution. In Ref.~\cite{Ghosh:2025rjh}, this length is replaced with $1/E_{\gamma} = 2\times10^{-14}$ cm. This replacement leads to a requirement on the photon density that is many orders of magnitude larger than the result given in Ref.~\cite{Diamond:2023scc}. As shown in Fig.~\ref{fig:fireball formation}, the entire region constrained by Ref.~\cite{Ghosh:2025rjh} is within the secondary fireball region for their assumed initial temperature of $T_i=337$ MeV. We emphasize that such a high value of temperature may not be realistic from the energetics of a $3M_\odot$ progenitor, but it is shown here just for comparison with the result of Ref.~\cite{Ghosh:2025rjh}. We also show in Fig.~\ref{fig:fireball formation} the secondary fireball formation region for a few other representative values of the temperature, along with the current astrophysical and laboratory constraints (shaded regions). We find that even for an sGRB with a core temperature of 20 MeV (which is a reasonable choice from Eq.~\eqref{eq: Temp_instanteneous}), a secondary fireball can form for ALP masses as large as 500 MeV.
\section{Fireball $X$-ray bounds}
\label{sec:diffuse bound}

\begin{figure}[t!]
    \centering
    \includegraphics[width=0.99\columnwidth]{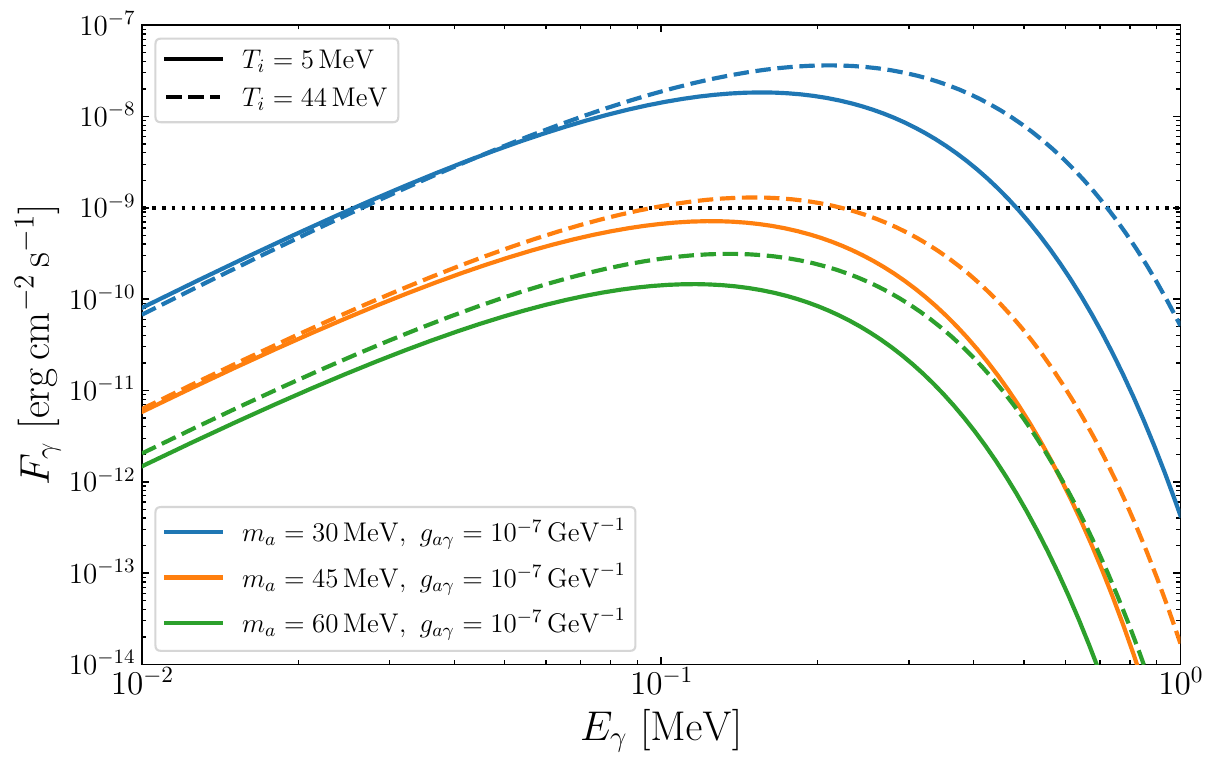}
    \captionsetup{justification=Justified}
    \caption{Photon flux at the Earth from a secondary ALP-induced fireball for a GRB located 100 Mpc away. The solid (dashed) curves are for $T_i=5~(44)$ MeV, while the blue, orange and green curves are for $m_a=30,$ 45, 60 MeV, respectively. We have fixed $g_{a\gamma}=10^{-7}~{\rm GeV}^{-1}$. The horizontal line shows the future sensitivity of $X$-ray telescopes. }
    \label{fig:photon spectra}
\end{figure}

Following Ref.~\cite{Diamond:2023cto}, we estimate the phase space distribution function of photon on the decoupling surface as
\begin{align}
    f(E_\gamma,\theta) = \dfrac{1}{\exp \left[ \dfrac{\gamma E_\gamma}{T_s} (1 - v \cos\theta) + \eta \right] - 1},
\end{align}
where $\eta = - \mu/T$. Thus, the number of photon emitted from the decoupling surface per unit time is given by
\begin{align}
    \frac{\mathrm{d}N_\gamma}{\mathrm{d}t} = 4 \pi r^2 \times 2 \int \frac{\mathrm{d}^3 p}{(2\pi)^3}f(E_\gamma,\theta),
\end{align}
where the prefactor $2$ takes into account two polarization states of the photon field. Since the emission from the secondary fireball is approximately isotropic, we are interested in the angle-integrated spectrum. The photon emission rate spectrum is thus given by
\begin{align}
    \frac{\mathrm{d}N_\gamma}{\mathrm{d}E_\gamma\mathrm{d}t} = 8 \pi r^2 \int_{-1}^1 \mathrm{d}y~\frac{2\pi E_\gamma^2 y}{(2\pi)^3}\dfrac{1}{\exp \left[ \dfrac{\gamma E_\gamma}{T_s} (1 - v y) + \eta \right] - 1}.
\end{align}
Here we defined $y \equiv \cos\theta$. The decoupling surface is likely to have velocity close to the speed of light ($v \simeq 1$), equivalently, $\gamma \gg 1$. Taking the limit $\gamma \to \infty$ while keeping in mind that $T \propto \gamma^{-1}$, and thus, $T \gamma \equiv \tau = \mathrm{const.}$, we have
\begin{align}
    \frac{\mathrm{d}N_\gamma}{\mathrm{d}E_\gamma\mathrm{d}t} = \frac{2}{\pi} r^2 E_\gamma^2 \int_{-1}^1 \mathrm{d}y\dfrac{y}{\exp \left[ \dfrac{E_\gamma y}{2 \tau} + \dfrac{E_\gamma \gamma^2}{\tau}(1-y) + \eta \right] - 1}.
\end{align}
This integral is analytically calculable. Up to the leading order~\cite{Diamond:2023cto},
\begin{align}
    \frac{\mathrm{d}N_\gamma}{\mathrm{d}E_\gamma\mathrm{d}t} = - \frac{2 r^2 E_\gamma \tau}{\pi\gamma^2} \log\left[ 1 - e^{-\eta - E_\gamma/(2\tau)} \right],
\end{align}
and thus, the flux ({\it i.e.}, the energy per unit area and time) at the Earth is
\begin{align}
    F_\gamma = - \frac{E_\gamma^3}{2 \pi^2 d_\mathrm{GRB}^2} \frac{r^2 \tau}{\gamma^2} \log\left[ 1 - e^{-\eta - E_\gamma/(2\tau)} \right],
\end{align}
where $d_\mathrm{GRB}$ is the distance to the GRB from the Earth. For reference, Fig.~\ref{fig:photon spectra} shows the photon flux observed at the Earth from a secondary fireball for a GRB located at distance $d_{\rm GRB}=100$ Mpc for different choices of the ALP mass (with a fixed $g_{a\gamma}$) and initial GRB temperatures. 

$X$-rays/$\gamma$-ray emission from this secondary fireball is isotropic. Thus, we could search for this signal from GRB sources whose GRB jets are not directed toward Earth. In that way, we avoid observing a signal from a secondary fireball on top of that from a GRB itself, making it easier to detect a secondary fireball-induced signal. 
The sensitivity of current space-based $X$-ray and gamma-ray instruments to sGRBs is typically at the level of $F_{\rm min} \sim 10^{-8}$--$10^{-7}\,\mathrm{erg\,cm^{-2}\,s^{-1}}$ on sub-second timescales, with variations driven by energy coverage and triggering methodology. The \textit{Neil Gehrels Swift Observatory}/BAT achieves a characteristic sensitivity of $\sim10^{-8}\,\mathrm{erg\,cm^{-2}\,s^{-1}}$ in the 15--150~keV band \cite{SwiftGRB2009ARA&A..47..567G}, while the \textit{Fermi} Gamma-ray Burst Monitor (GBM), with its much larger instantaneous sky coverage, has a higher effective threshold of $\sim3\times10^{-8}\,\mathrm{erg\,cm^{-2}\,s^{-1}}$ in the 50--300~keV range \cite{GBM_Meegan2009}. Recently launched \textit{SVOM}/ECLAIR instrument has a similar sensitivity in the 4--150~keV range \cite{SVOM2022IJMPD..3130008A}. Future $X$-ray missions such as \textit{THESEUS} \cite{THESEUS2018} and GECCO \cite{GECCO_Orlando:2021get} are expected to reach sensitivities up to an order of magnitude better than \textit{Swift}/BAT. Similarly, next-generation MeV gamma-ray observatories, such as \textit{AMEGO} \cite{AMEGO:2019gny} and \textit{e-ASTROGAM} \cite{e-ASTROGAM:2017pxr}, are designed to achieve a flux sensitivity of $F_{\rm min} \sim few \times 10^{-9}\,\mathrm{erg\,cm^{-2}\,s^{-1}}$ in the $\sim0.2$--$100$~MeV range. Figure~\ref{fig:photon spectra} shows an optimistic flux sensitivity of future mission to sGRBs at $F_{\rm min} = 10^{-9}\,\mathrm{erg\,cm^{-2}\,s^{-1}}$ across the $X$-ray and MeV gamma-ray bands. The emission from ALP-induced secondary fireball is not background free, however, as short-duration transients such as fast $X$-ray transients, low-luminosity GRBs and tidal disruption events are known to emit in the same $X$-ray range. We rather hope to constrain the ALP parameters from non-detection of the secondary fireball signature in these transients. In other words, assuming a non-observation of excess $X$-ray emission in these instruments, we can rule out the ALP parameter space that produces a flux in excess of $F_{\rm min}$ (the curves going above the horizontal line in Fig.~\ref{fig:photon spectra}). For instance, for $g_{a\gamma}=10^{-7}~{\rm GeV}^{-1}$, a 30 MeV ALP can be ruled out for both choices of the initial GRB temperature, while a 45 MeV ALP can be ruled out only for the aggressive choice of $T_i$, and the 60 MeV ALP cannot be ruled out for either choice of $T_i$.

\begin{figure}[t!]
    \centering
    \includegraphics[width=0.99\columnwidth]{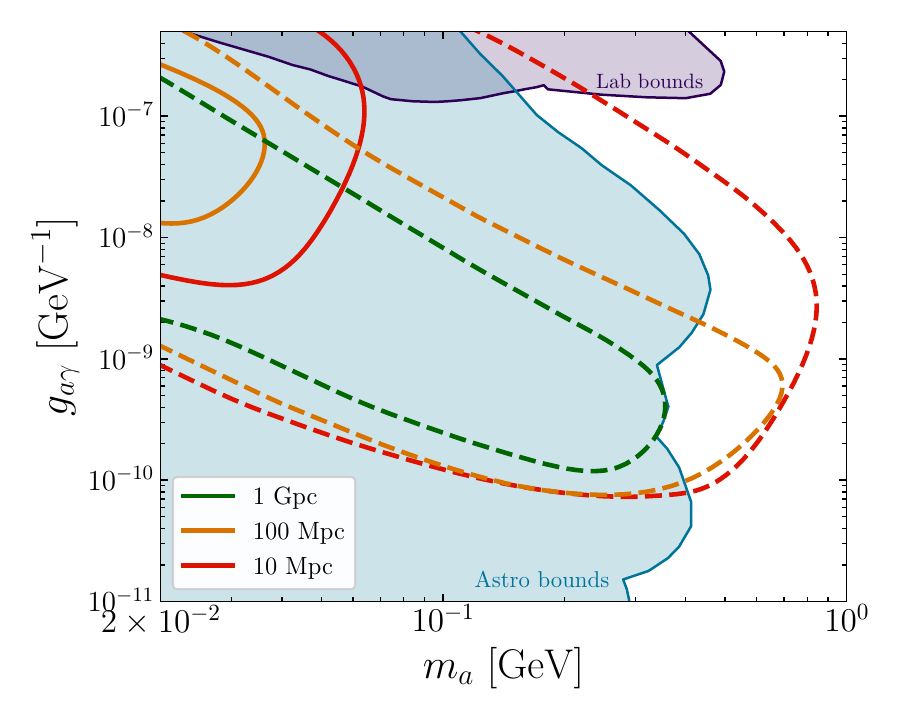}
    \captionsetup{justification=Justified}
    \caption{Sensitivity of future $X$-ray observations to the photon flux from secondary ALP-induced fireball in sGRBs. The solid (dashed) contours are for an initial fireball temperature of $T_i=5~(44)$ MeV, while the green, orange and red contours are for GRB distance scale of 1 Gpc, 100 Mpc and 10 Mpc, respectively.  The current lab bounds (purple shaded) and astrophysical bounds (cyan shaded) are also shown for comparison. }
    \label{fig: fireball bounds}
\end{figure}

Figure~\ref{fig: fireball bounds} exhibits the region in the ALP parameter space that can thus be excluded by future $X$-ray observations. Here, we have taken two different benchmark values for the initial fireball temperature: $T_i=5$ MeV (solid) and 44 MeV (dashed), and three different values for the distance to the sGRB: $d_{\rm GRB}=1$ Gpc (green), 100 Mpc (orange) and 10 Mpc (red).\footnote{The local rate of GRBs is $\approx 5$~Gpc$^{-3}$~yr$^{-1}$, while Swift/BAT sGRBs give an upper limit for the all-sky rate of $<4~{\rm yr}^{-1}$ at $d< 200$ Mpc~\cite{Mandhai:2018cdl,Dichiara:2019kuw, Mandel:2021smh}.} For comparison, we also show the laboratory bounds (purple shaded) and other astrophysical bounds (cyan shaded), taken from the compilation of Ref.~\cite{AxionLimits}. 
We note that cosmological observables, such as $\Delta N_\mathrm{eff}$, the primordial abundance of nuclei, and spectral distortion of cosmic microwave background, are sensitive to the production of ALPs in the early universe~\cite{Cadamuro:2010cz, Escudero:2025avx}. These cosmological bounds are not shown here as they depend on the reheating temperature and, more importantly, they are insensitive to $m_a \gtrsim 10~\mathrm{MeV}$ ALPs if the reheating temperature takes the lowest possible value ($T_\mathrm{RH} \simeq 5~\mathrm{MeV}$) with instantaneous reheating~\cite{Depta:2020wmr}. 

From Fig.~\ref{fig: fireball bounds}, we find that non-observation of excess transient photon flux, uncorrelated with known $X$-ray sources, over the diffuse $X$-ray background expected in future $X$-ray observations  might be able to constrain new ALP parameter space, provided there is an exceptionally energetic GRB within $\sim$100 Mpc. Nevertheless, the ALP-induced secondary fireball formation in GRBs provides a new probe of ALPs, irrespective of the fate of the progenitor and irrespective of the alignment of the GRB jet.  

\section{Conclusions}
\label{sec: conclusions}

In this paper, we carefully estimated ALP production inside a GRB jet by considering two different assumptions about the burst launching mechanism. More commonly, it is assumed that the GRB is powered by the relativistic outflow that persists for a duration comparable to the burst, leading to the conservative estimation of the GRB fireball temperature. In an alternate scenario, we may assume that all of GRB energy is instantaneously injected to a fireball of a radius slightly larger than the Schwarzschild radius of a remnant. We find that, in both cases, the GRB fireball temperature is unlikely to amount to $T_i$ above ${\cal O}(10)$ MeV for both short and long GRBs\footnote{We have checked that long GRBs typically result in lower $T_i$, leading to weaker constraints on ALPs. Thus, we focus only on short GRBs in this work.} as long as we consider typical GRB energy or luminosity. This implies that ALPs are produced inside the jet less efficiently than previously expected, and thus, the cooling bounds on the axion-photon coupling are altered. We recalculate the bounds for several choices of core temperature, and show that these bounds are unlikely to outperform existing astrophysical bounds unless there exist extremely energetic GRB events. This result is summarized in Fig.~\ref{fig:energylosscontours}. 

In addition, we also investigated the formation of a secondary fireball in this scenario. We find that photons from the decay of ALPs that are produced inside the primary GRB fireball could form an expanding shell of thermalized photons, electrons, and positrons, emitting $X$-rays from its surface. This secondary fireball turns out to be formed in a wide region in the $m_a$-$g_{a\gamma}$ parameter space as shown in Figs.~\ref{fig:fireball temperature} and \ref{fig:fireball formation}. 

We estimated the ensuing $X$-ray flux at Earth and its observability by current and future $X$-ray telescopes (see Fig.~\ref{fig:photon spectra}). In the absence of an excess $X$-ray flux, we can thus set new bounds on the ALP parameter space, as shown in Fig.~\ref{fig: fireball bounds}. We find that these results can be comparable to or better than current laboratory and astrophysical constraints on ALPs, provided there is an exceptionally energetic GRB event within $\sim 100$ Mpc.  

Lastly, we note that our argument involves several subtleties. Firstly, our estimation of the ALP production rate is approximate mainly because we assume (i) no recoil of a target for the Primakoff process even though the GRB fireball temperature is higher than the mass of main scatterer (electrons) and (ii) presence only of two transverse modes with the dispersion relation $\omega^2 = k^2 + \omega_\mathrm{pl}^2$. The ALP production rate is modified if these assumptions are removed, as discussed {\it e.g.}, in Ref.~\cite{Altherr:1990wi,Altherr:1992mf,Altherr:1993zd}. We also treat a secondary fireball as a (at least nearly) spherically-symmetric object in the laboratory frame despite that the GRB fireball, where ALPs are produced, has nonzero bulk Lorentz factor $\gamma = r/r_i$. While most of the ALPs are produced in the innermost region where $\gamma$ is small, a small fraction of ALPs produced in the outer region possibly alter our discussion on a fireball formation and estimation of its $X$-ray signal. We also remark on several practical considerations in our setup. It may be necessary to distinguish a secondary fireball-induced $X$-ray signal from that of an astrophysical object, in particular, if a GRB jet is not directed toward Earth. Rarity of the Galactic GRB event might present an additional challenge for our approach.

\acknowledgments
 We thank Oindrila Ghosh and Tim Linden for correspondence regarding their work~\cite{Ghosh:2025rjh}. We also thank Jim Buckley and Manel Errando for  discussions on GRB properties, Melissa Diamond and Damiano Fiorillo for discussions on fireball formation, and Irene Tamborra for helpful comments. We would like to thank The Institute for Underground Science at Sanford Underground Research Facility (SURF) for local hospitality during PPC 2025 where a crucial part of this work was done.  C.V.C. was generously supported by Washington University in St. Louis through the Edwin Thompson Jaynes
Postdoctoral Fellowship. S.D. was supported by a McDonnell Center for the Space Sciences Postdoctoral Fellowship. The work of B.D. and T.O. was partly supported
by the US Department of Energy under grant No. DE-SC0017987. B.D. was also partly supported by a Humboldt Fellowship from the Alexander von Humboldt Foundation. S.R.
was partially supported by a National Research Foundation (NRF) of South Africa grant facilitated through the National Institute of Theoretical and Computational Sciences (NITheCS).
\appendix
\section{Comments on the Previous Work~\cite{Ghosh:2025rjh}}
\label{sec:previous}
The present work was largely inspired by Ref.~\cite{Ghosh:2025rjh}, which argued that strong bounds could be placed on GeV-scale ALPs based on cooling of sGRBs by ALP emission. However, we find several of the assumptions made in that work difficult to justify, particularly regarding the initial radius and temperature of the fireball (in this Section, the term ``fireball" always refers to the initial, Standard Model fireball of the GRB, not the secondary fireball produced by ALPs). In this Section, we discuss various sources of uncertainty in the ALP production rate, and compare the assumptions of Ref.~\cite{Ghosh:2025rjh} with values that are more typical in the literature, or easier to justify.

\subsection{Trapping via Decay}

In Ref.~\cite{Ghosh:2025rjh}, the probability for an ALP to escape the fireball is encoded in a factor
\begin{equation}
    e^{-r/\lambda_{a\rightarrow \gamma \gamma}}\,,
\end{equation}
with $r$ being the distance from the center of the gravitational remnant. This factor suggests that ALPs produced close to $r = 0$ are more likely to escape without decaying than ALPs produced at the outer edge of the fireball. We instead use a factor

\begin{equation}
    e^{-\left(\frac{r_{\rm esc} - r}{\lambda_{a\rightarrow \gamma \gamma}}\right)}\,,
\end{equation}
where $r_{\rm esc}$ is the distance from the center of the gravitational remnant at which the fireball becomes optically thin. Our formula results in a weaker bound at large masses, because it accounts for the large distance that heavy ALPs, which will be most efficiently produced at small radius, must travel in order to escape the fireball. However, we note that this difference does not imply a flaw in Ref.~\cite{Ghosh:2025rjh}. The analysis in Ref.~\cite{Ghosh:2025rjh} focuses on ALP production and emission at very early times, when the fireball has not yet expanded to the radius $r = r_\mathrm{esc}$. The optically thick region is smaller in this regime, corresponding to a shorter distance for ALPs to escape from such region. In fact, the authors of Ref.~\cite{Ghosh:2025rjh} have checked that the resulting impact of the different escape criteria is modest.

\subsection{Consistency of Energy and Luminosity Values}

Ref.~\cite{Ghosh:2025rjh} sets limits on heavy ALPs by requiring that the energy loss rate due to ALP production not exceed the intrinsic GRB luminosity. Specifically, the authors require that the ALP luminosity not exceed physical GRB luminosity of $10^{50}$ erg/s, corresponding to an isotropic-equivalent luminosity of approximately $10^{52}$ erg/s, given their choice of $\Delta \theta$. However, this is inconsistent with the energy release required to produce the assumed temperatures. Again, we note that an isotropic-equivalent energy of order $10^{55}$ erg is required in order to produce the temperatures assumed in Ref.~\cite{Ghosh:2025rjh} (see the text following Eq.~\eqref{eq: Temp_instanteneous}). However, given that sGRBs have a duration of at most 2 s, this implies a minimum isotropic-equivalent GRB luminosity of $5 \times 10^{54}$ erg/s, two to three orders of magnitude larger than the luminosity criterion used in Ref.~\cite{Ghosh:2025rjh}. This means that ALPs with parameters that sit exactly along the limit reported in Ref.~\cite{Ghosh:2025rjh} would have a percent- or sub-percent-level effect on the energetics of the $10^{55}$ erg GRB under consideration.

\subsection{Assumed GRB Energy}

As stated above, even under the assumption that $r_i = r_s$, and even using the same formula for temperature used in Ref.~\cite{Ghosh:2025rjh}, reaching temperatures of a few hundred MeV for a $3M_{\odot}$ merger remnant requires an isotropic-equivalent energy of around $10^{55}$ erg. However, the most energetic known sGRB was estimated to have an isotropic-equivalent energy of $2.1 \times 10^{53}$ erg~\cite{deUgartePostigo:2006ub}, assuming a redshift of $z = 4.6$ (a different estimate, $z = 1.7$, yields a total energy of $2.9 \times 10^{52}$ erg). Using $2.1 \times 10^{53}$ erg as the isotropic-equivalent energy and a 3$M_{\odot}$ remnant mass yields a temperature of 99 MeV, a factor of a few lower than the temperature quoted in Ref.~\cite{Ghosh:2025rjh}, when using the same assumptions.

\subsection{Initial Launch Radius}\label{subsection:ri}

In Ref.~\cite{Ghosh:2025rjh}, the initial radius of the fireball is taken to be equal to $r_s$, the Schwarzschild Radius of the remnant sourcing the fireball. However, we find this choice very aggressive. 
For a simple comparison, it is much smaller than the values given for the initial radius in classic GRB literature~\cite{Piran:1999kx}, the values often chosen in simulations of GRBs~\cite{MacFadyen:1998vz}, and those previously used in setting limits on Beyond-Standard-Model particles~\cite{Tu:2015lwv}. We therefore take a value of 3$r_s$ as the launch radius of the fireball in our work, a value which is much more typical in the literature and coincides with the innermost stable circular orbit.




\begin{table*}[htb!]
\begin{center}
\begin{tabular}{|c|c|c|c|c|}
    \hline
    Benchmark Point & Difference from Ref.~\cite{Ghosh:2025rjh} & Core Temperature [MeV] & $\edit{E_\mathrm{iso}}$ [erg] & Luminosity Limit [erg/s] \\
    \hline\hline
    1 & Ref.~\cite{Ghosh:2025rjh} & 337 & 2.8 $\times$ 10$^{55}$ & 10$^{50}$ \\
    2 & Decay Trapping and Primakoff Production & 337 & 2.8 $\times$ 10$^{55}$ & 10$^{50}$ \\
    3 & Consistency of Luminosity Limit & 337 & 2.8 $\times$ 10$^{55}$ & 1.4 $\times$ 10$^{53}$ \\
    4 & Lower Isotropic Energy & 99 & 2.1 $\times$ 10$^{53}$ & 10$^{51}$ \\
    5 & $r_i = 3r_s$ & 44 & 2.1 $\times$ 10$^{53}$ & 10$^{51}$ \\
    6 & Luminosity Formula [Eq.~\eqref{eq:luminositytemperature}] & 5 & 2.1 $\times$ 10$^{53}$ & 10$^{51}$ \\
    \hline
\end{tabular}
\end{center}
\captionsetup{justification=Justified}
\caption{Various choices of core temperature, total energy, and limiting luminosity discussed in this Section. Case 1 is the result of Ref.~\cite{Ghosh:2025rjh}. Case 2 is our recomputation of their result, after correcting typos in the escape criteria, and including Primakoff production. Case 3 further requires that the luminosity limit be equal to $\frac{\edit{E_\mathrm{iso}}}{2\Delta\theta^2} \frac{1}{\textrm{s}}$. Case 4 requires that the energy of the sGRB not be larger than the energy of the most energetic observed sGRB. Case 5 requires that $r_i = 3r_s$. Case 6 uses Eq.~\eqref{eq:luminositytemperature} instead of Eq.~\eqref{eq: Temp_instanteneous} to compute the GRB temperature. \label{tab:GRB Parameters}}
\end{table*}
\begin{figure}[t!]
    \centering
    \includegraphics[width=0.99\columnwidth]{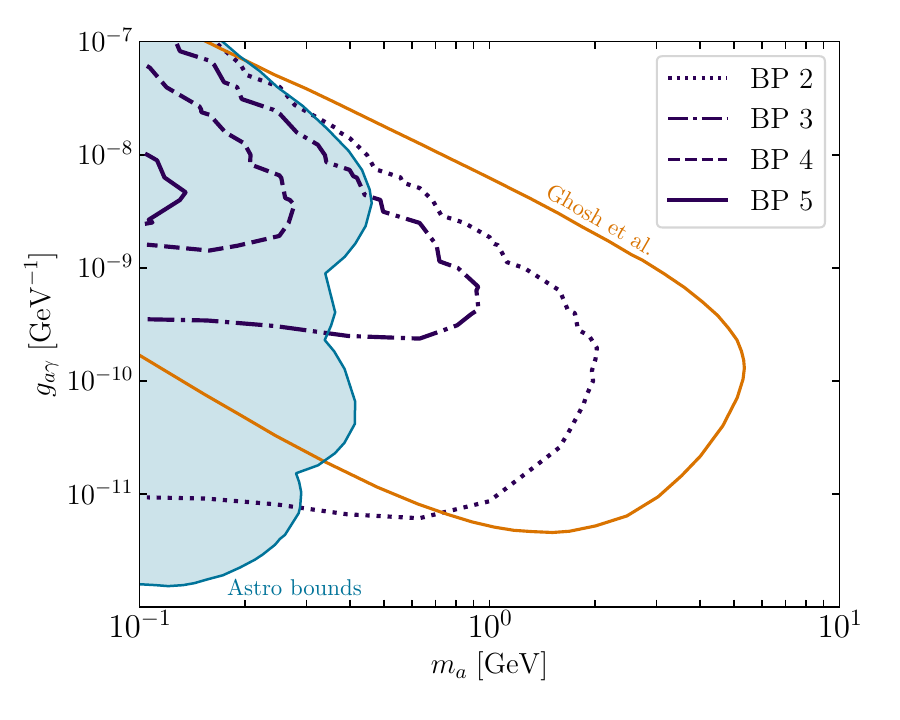} \captionsetup{justification=Justified}
    \caption{Constraints on ALP parameters based on sGRB energy loss, assuming the different combinations of sGRB parameters presented in Table~\ref{tab:GRB Parameters}. Note that for Case 6, the exclusion region only extends up to a mass of around an MeV, and thus does not appear on this plot.}
    \label{fig:energylosscontours}
\end{figure}

\subsection{Temperature Formula}

Ref.~\cite{Ghosh:2025rjh} follows Refs.~\cite{Tu:2015lwv,Koers:2005ya} in computing the temperature of the fireball in terms of the energy released, using Eq.~\eqref{eq: Temp_instanteneous}. This formula may be derived by assuming that all of the GRB's energy is deposited instantaneously in a spherical region of radius $r_i$. The assumption of instantaneous energy injection here can lead to extremely high temperatures: assuming $r_i = r_s$, Ref.~\cite{Ghosh:2025rjh} computes temperatures of several hundred MeV for sGRBs with various remnant masses. For reference, we find that an isotropic-equivalent energy $\edit{E_\mathrm{iso}}$ of order $10^{55}$ erg is required to produce a temperature of 261 MeV using this formula. Refs.~\cite{Tu:2015lwv,Koers:2005ya} compute a temperature of 18 MeV using a more conservative value of $\edit{E_\mathrm{iso}}$.

However, an alternative to instantaneous injection and a more commonly accepted scenario is that GRBs are powered by a sustained, relativistic outflow lasting approximately the duration of the burst. In this picture, the temperature $T_i$ is given in terms of the luminosity $L$ by Eq.~\eqref{eq:luminositytemperature}.
This formula results in a much lower temperature for the same energy: for an sGRB of energy $10^{55}$ erg and duration 1 s, such that $\edit{L_\mathrm{iso}} = \edit{E_\mathrm{iso}}/\textrm{s}$, we obtain a temperature of 22.7 MeV, an order of magnitude lower than the temperature computed for the same energy by assuming instantaneous injection. This particular issue has been raised already in Ref.~\cite{Candon:2025ypl}.

Figure~\ref{fig:energylosscontours} summarizes the variation in ALP constraints coming from these considerations. For more physically realistic benchmark cases 5 and 6, we find that the energy loss constraint is far weaker than the existing astrophysical constraints and does not cover any new ALP parameter space, unlike the claim made in Ref.~\cite{Ghosh:2025rjh}.

\bibliographystyle{utcaps_mod}
\bibliography{bibliography}
\end{document}